**POSITION PAPER**

**EU DIGITAL REGULATION AND GUATEMALA:**
**AI, 5G, AND CYBERSECURITY**

VICTOR LÓPEZ JUÁREZ

MARCH 2024

## I. EU DIGITAL GOVERNANCE: MAPPING THE REGULATORY CORE

The European Union (hereinafter 'EU') is a *sui generis*[1] organization that aims, *inter alia*, to safeguard fundamental rights[2] and foster the internal market among its Member States.[3]

With the rapid growth of new technologies and connectivity, the EU is invoking digital sovereignty[4] to protect its core values, rights, and principles against the challenges of the global digital society.[5]

In response to those challenges, Europe has implemented different measures as part of its digital strategy. The Data Protection Directive was adopted in 1995. The E-Commerce Directive has been in place since 2000 and harmonizes the rules for online services. More recently, the first digital agenda[6] was adopted from 2010 to 2020, focused on improving access to digital goods and services.

Later, the EU launched a second digital agenda[7], spanning from 2020 to 2030, along with the Digital Decade Policy Programme for 2030.[8] These initiatives are designed to address the significant changes brought by digital transformation and Europe's geopolitical aspirations.

---

[1] Wolfram Kaiser et al, *The History of the European Union* (Routledge 2009) 4.
[2] European Union, Treaty of Lisbon Amending the Treaty on European Union and the Treaty Establishing the European Community, 13 December 2007, 2007/C 306/01 art 3(3).
[3] ibid, art 3(4).
[4] Julia Pohle and Thorsten Thiel, 'Digital Sovereignty' (2020) 9 Internet Policy Review 19, 2.
[5] Edoardo Celeste, 'Digital Sovereignty in the EU: Challenges and Future Perspectives', in Fabbrini, F., Celeste, E. & Quinn, J. (Eds.), *Data Protection Beyond Borders: Transatlantic Perspectives on Extraterritoriality and Sovereignty* (Oxford Hart Publishing, 2020) 220.
[6] See European Commission, '*A Digital Agenda for Europe'* COM(2010)245, 3.
[7] See European Commission, '*Shaping Europe's Digital Future'* (Publications Office 2020) 3.
[8] Decision (EU) 2022/2481 of the European Parliament '*Digital Decade Policy Programme 2030'* (2022).

# A. ARTIFICIAL INTELLIGENCE: THE AI ACT AS TRANSNATIONAL GOVERNANCE

Artificial Intelligence is a machine-based system that can work autonomously, generate content, and make decisions based on input.[9] AI is becoming increasingly prevalent in our daily lives, with numerous applications and use cases.[10]

The EU has launched an AI strategy to become a world-class hub for AI while ensuring that it is human-centric and trustworthy. The cornerstone of this strategy is the AI Act, which is the first-ever legal framework on AI that addresses its potential risks and promotes its responsible use.

The AI Act imposes precise requirements and obligations on AI developers and deployers, addresses four levels of risks created by AI, and implements rigorous measures to mitigate those risks. Additionally, the AI Act establishes a governance structure at the European and national levels. The European AI Office will supervise the enforcement and implementation of the AI Act within the Member States.

In addition to the AI Act, the EU's AI strategy includes complementary initiatives such as the AI Innovation Package and the Coordinated Plan on AI. These measures are designed to increase Europe's global competitiveness. By setting up a comprehensive AI digital policy, the EU retains its first-mover advantage[11] and has established itself as a global leader in shaping AI.

---

[9] AI Act, Regulation No 1234/2023, OJ L 123/45 2023 art 3(1).

[10] McKinsey Global Institute, 'Notes from the AI Frontier: Insights from Hundreds of Use Cases. Discussion Paper' (2018) 30.

[11] Steven Feldstein, 'Evaluating Europe's Push to Enact AI Regulations: How Will This Influence Global Norms?' (2023) Democratization, 3.

## B. 5G CONNECTIVITY: FROM SECURITY TOOLBOX TO TRUST ARCHITECTURE

5G is a term used to describe the latest generation of telecommunication systems that offers significant enhancements in terms of speed and overall performance to meet the growing demand for connectivity.[12] Compared to its predecessor, 5G is a hundred times faster[13] at data processing. Its highest impact is expected in essential sectors such as healthcare, transportation, and manufacturing.

The EU recognized the need for a 5G policy[14] and adopted various strategic initiatives and regulations,[15] including the 5G Action Plan in 2016 to boost the deployment of 5G infrastructure and services across Europe and to make it a reality for all citizens and businesses.

The EU's 5G Action Plan includes measures such as coordinated deployment, spectrum availability, urban deployment, innovation facilitation, and investment.

In 2020, the EU implemented the 5G Toolbox to guarantee the security and integrity of 5G networks as critical infrastructure. The toolbox provides a comprehensive set of security measures aimed at addressing potential risks and ensuring the secure deployment of 5G networks. It outlines detailed mitigation strategies for each identified risk and recommends a set of critical strategic and technical measures to be taken by Member States.[16]  Finally, it is essential to note that the EU is planning to update its 5G action plan, which includes plans for 6G.

---

[12] European Commission, '5G Networks and Technology' 1.
[13] Naser Al-Falahy and Omar Y Alani, 'Technologies for 5G Networks: Challenges and Opportunities' (2017) 19 IT Professional 12, 18.
[14] Berna Akcali Gur, 'Cybersecurity, European Digital Sovereignty and the 5G Rollout Crisis' (2022) 46 Computer Law & Security Review 105736, 3.
[15] See Directive (EU) 2018/1972 of the European Parliament and of the Council of 11 December 2018.
[16] European Commission, '*EU Toolbox for 5G Security*' (Publications Office 2020) 2.

## C. CYBERSECURITY: ENISA, NIS2, AND THE DUTY OF DIGITAL CARE

Cybersecurity involves protecting networks, information systems, and users[17] from cyber threats that can cause harm or disruption.[18]  The European Union has developed digital policies and a cybersecurity strategy to enhance resilience to cyber threats, secure data, and protect citizens' rights.

This strategy aims to create collective capabilities to respond to significant cyberattacks while outlining plans to work with international partners to ensure security and stability in cyberspace.

The European Union Agency for Cybersecurity (ENISA) was created to safeguard the EU's cybersecurity. It provides support to Member States, EU institutions, and businesses in various vital areas, such as the implementation of the two Directives on the Security of Network and Information Systems, which ensure that strong government bodies are established to supervise cybersecurity within their respective countries.

The EU has implemented the Cybersecurity Act, which strengthens ENISA's role and establishes a cybersecurity certification framework for products, processes, and services.  Additionally, the EU's Cyber Resilience Act sets regulations for cybersecurity requirements in products with digital components. Through a combination of policies, strategies, and dedicated agencies like ENISA, the EU aims to reinforce cybersecurity.

---

[17] Cybersecurity Act, Regulation (EU) 2019/881 of the European Parliament and of the Council, (2019) OJ L 156/1, art 2(1).
[18] Ibid, art 2(8).

## II. ACTUAL AND POTENTIAL EFFECTS IN GUATEMALA

### A. AI: THE BRUSSELS EFFECT, ALGORITHMIC SOVEREIGNTY, AND POLICY SPACE

The EU's digital policy on AI aims to establish uniform regulations in the field. While these regulations are intended to promote trust and innovation within the EU, their repercussions can reach far beyond European borders.

One such repercussion is Europe's regulatory imperialism, which is the EU's conscious effort to impose its regulations over other jurisdictions, often motivated by protectionism and the desire to impose standards on a global scale.[19]   Europe's exercise of regulatory imperialism is evident in its imposition of stringent regulations that not only govern its internal market but also have extraterritorial implications for third countries outside the EU.

Another repercussion of EU regulations is the Brussels effect, in which third countries and companies comply with EU regulations voluntarily without the need for direct intervention. The Brussels effect is considered an unintended by-product of the EU regulatory power.[20] Countries like Guatemala may require substantial investments in regulatory alignment, technology, and human resources to comply *de facto* or *de jure* with the EU's AI regulations.

The EU's AI regulations could also create barriers to trade for Guatemalan businesses seeking access to the European market. Guatemalan companies exporting AI products to the EU must comply with the obligations set out in the AI regulations and assess the conformity of their products.  Failure to meet EU standards could result in restrictions on the export of AI-related applications and services to the EU, limiting Guatemala's economic opportunities and restraining innovation in the AI sector.

---

[19] Anu Bradford, 'Exporting Standards: The Externalization of the EU's Regulatory Power via Markets' (2015) 42 International Review of Law and Economics 158, 159.
[20] Anu Bradford, *The Brussels Effect: How the European Union Rules the World* (Oxford University press 2020) 28.

If Guatemalan companies do not want to forgo the entire EU market, they will have to comply with the EU's AI regulations, leading to a *de facto* Brussels effect, and in the future, those companies will most likely lobby the Guatemalan government to implement the same standards nationally to level the playing field for all competitors, resulting in a *de jure* Brussels effect.

The EU's AI regulatory power can also negatively affect Guatemalan democracy since unelected European officials impose regulations affecting Guatemala without direct democratic oversight or accountability. External regulations may not always be in the best interest of Guatemalan welfare and could potentially lead to strained international relations.

Adhering to the EU's AI regulations might also diminish Guatemalan sovereignty since the country does not have any direct input or say in the decision-making process, challenging its authority and independence. Additionally, the EU's regulatory approach to AI may set a precedent for the rest of the jurisdictions. This hegemonic influence could exacerbate the digital gap between rich and poor countries, further marginalizing countries like Guatemala in the global AI landscape.

While the EU's AI regulations aim to promote human-centric and trustworthy AI technologies, their extraterritorial implications have the potential to perpetuate the Brussels effect, regulatory imperialism, and colonialist phenomena[21] particularly for Guatemala.

---

[21] Edoardo Celeste, 'Brexit and the Risks of Digital Sovereignism' (2024) Brexit Institute Working Paper Series No 01/2024, 196.

## B. 5G: ENVIRONMENTAL TRADEOFFS, INFRASTRUCTURE RIGHTS, AND SOCIAL LICENSE TO OPERATE

The EU's 5G digital policy promises remarkable advancements in connectivity. However, alongside these anticipated benefits, significant environmental concerns arise, specifically for Guatemala.  One pressing issue is the anticipated increase in energy consumption associated with the implementation of 5G infrastructure[22]. As Guatemala and its companies navigate the deployment of new infrastructure, such as cells and antennas, there's a potential for heightened energy demands, which, if not managed effectively, could lead to a greater carbon footprint and environmental impact.

Additionally, the accelerated turnover of electronic devices and the introduction of 5G networks may exacerbate electronic waste issues, posing environmental risks if appropriate waste management is not in place. Moreover, the production of 5G infrastructure components requires raw materials, including rare earth metals, often extracted in environmentally sensitive areas. This extraction process could result in water pollution and other adverse environmental consequences.

Furthermore, concerns regarding electromagnetic radiation from 5G infrastructure, although operating within established safety limits, may raise valid concerns about public health impacts for Guatemala.

Finally, it is highly probable that these negative impacts on the environment would lead to side effects that violate Human Rights.  This is because the environment and fundamental rights are closely related. The European Court of Human Rights (ECHR) has recognized this connection and has declared violations to the right to life,[23] health,[24] private and family life[25] due to environmental pollution.[26]

---

[22] Céline Perea, 'Digital Sobriety: From Awareness of the Negative Impacts of IT Usages to Degrowth Technology at Work' (2023) 3.
[23] *Öneryildiz v Turkey* App No 48939/99 (ECHR, 30 November 2004).
[24] *Bacila v Romania* App No 19234/04 (ECHR, 30 March 2010).
[25] *Borysiewicz v Poland* App No 71146/01 (ECHR, 1 July 2008) .
[26] See Francesca Magistro, 'Le droit à un environnement sain revisité' (Schulthess éditions romandes 2017) 97.

### C. CYBERSECURITY: COMPLIANCE FRICTIONS, CAPACITY GAPS, AND MARKET STRUCTURE

Guatemala faces significant economic challenges in complying with the EU's Cybersecurity policy. The first challenge is the cost of implementation. Developing and maintaining robust cybersecurity measures as outlined by the EU requires substantial investments, which can hinder overall financial sustainability.

Another challenge is the additional compliance costs[27] associated with EU regulations. Guatemalan businesses willing to do business in the EU may face extra financial burdens associated with audits, reporting, and possible fines. These costs can disproportionately affect smaller companies with limited resources and discourage businesses from formalizing, hindering economic growth.

Guatemala may also potentially face a shortage of cybersecurity professionals.[28] The high demand for qualified personnel often outstrips supply, driving up talent acquisition and retention costs. This lack of skilled professionals can hinder Guatemala's ability to implement and manage its cybersecurity strategy effectively, forcing it to rely on expensive foreign expertise.

Summarizing, Guatemala must carefully balance its need to protect its digital infrastructure from cyber threats with ensuring the economic sustainability of its businesses. Finding innovative solutions and seeking international support to mitigate these economic burdens is crucial for Guatemala to safeguard its digital future without compromising its economic growth.

---

[27] Shari Pfleeger, 'Cybersecurity Economic Issues: Clearing the Path to Good Practice' (2008) 35.
[28] Karen Grass, 'Die Debatte Zu Digitalisierung Und Arbeitsmarkt in Europa' (2016) 9.

### III. GUATEMALA'S LEGAL POSTURE AND THE REFORM TRACK

#### A. ARTIFICIAL INTELLIGENCE: BUILDING A COHERENT NATIONAL FRAMEWORK UNDER THE SANTIAGO DECLARATION

Guatemala does not yet have specific legislation for AI. However, the Santiago Declaration to promote Artificial Intelligence in Latin America and the Caribbean,[29] adopted in October 2023 and signed by 20 more countries, provides valuable guiding principles for Guatemala's approach to AI policy. The declaration emphasizes the importance of human-centered, inclusive, transparent, robust, and accountable AI. These principles underscore the need for AI development and deployment to respect human rights, promote inclusivity, ensure transparency and reliability, and hold those involved in the AI lifecycle accountable.

#### B. 5G: SPECTRUM GOVERNANCE, NETWORK SECURITY, AND THE ROAD TO 6G READINESS

In 2023, Guatemala passed a reform to Article 65 of the General Telecommunications Law,[30] allocating funds to support the National Digital Connectivity Plan.[31] The National strategy aims to facilitate the deployment and adoption of 5G technology across the country while ensuring a transparent and competitive process for future spectrum allocations. This will involve expanding the available spectrum bands beyond the initial 2.5 GHz and encourage participation from a wider range of telecommunication companies. Increased competition will drive

---

[29] See Santiago Declaration to promote Artificial Intelligence in Latin America and the Caribbean 2023 (CL).
[30] See Congress, General Telecommunications Law 1996 (GT).
[31] See Communications Ministry, Plan Nacional de Conectividad y Banda Ancha 2018 (GT).

down service costs, making 5G more accessible to Guatemalan people and companies.

## C. CYBERSECURITY: NATIONAL STRATEGY, INSTITUTIONAL DESIGN, AND MINIMUM SECURITY BASELINES

In 2018, Guatemala took a significant step towards online security by implementing a National Cybersecurity Strategy.[32] This strategy aims to safeguard citizens' rights, assets, and personal data within the digital landscape.  The approach further underlines the concept of shared responsibility, highlighting its role in strengthening the nation's resilience against cyber threats. To facilitate this collaboration and inter-sectoral coordination, a National Committee on Cybersecurity was established. However, Guatemala has faced challenges in implementing this cybersecurity strategy due to a lack of political will, financial resources, and infrastructure.

## IV. FIVE GUARDRAILS FOR RESPONSIBLE EU RULEMAKING WITH THIRD COUNTRY IMPACTS

Guided by well-established practices[33] in policy analysis[34] Guatemala proposes the following five responsible rule-making principles:

### 1. Digital Constitutionalism to moderate Regulatory Power.

Digital constitutionalism can be defined as the need for evolving legal frameworks, rights, and principles to address the challenges and opportunities presented by digital transformation.[35] Digital Constitutionalism can be used as a conceptual barrier to digital sovereignism[36] for the benefit of third countries. Therefore, by promoting digital constitutionalism, the EU may resist the urge to pursue complete sovereignty over digital policies and instead focus on upholding universal rights and principles in the digital space. This approach will allow the EU to exercise its regulatory power moderately, implementing more equitable digital policies for the benefit of third countries.

---

[32] See Interior Ministry, 'Guatemalan National Cybersecurity Strategy' (2018).
[33] See European Union ETF, *Guide to Policy Analysis* (Publications Office 2018) 37.
[34] See Eóin Young, *Writing Effective Public Policy Papers* (Open Society Institute 2002) 18.

## 2. Green IT to counter environmental impact.

The EU can reduce the environmental footprint of its digital policies by implementing Green IT practices. These include mandatory ecological impact assessments for new IT products, eco-friendly design standards for hardware, and holding manufacturers responsible for their products' lifecycle (disposal and recycling).

Furthermore, promoting circular economy practices like product reuse and refurbishment can minimize waste. Finally, incentivizing green innovation in the IT sector can accelerate the development of sustainable technologies. These measures can pave the way for more responsible, environmentally friendly digital policies within the EU and third countries.

## 3. Regulatory Impact Assessment to anticipate economic burden.

Regulatory Impact Assessment[37] (RIA) is a valuable tool for understanding the potential economic effects of regulations and promoting evidence-based decision-making. Evaluating the cost-benefit of EU digital policies beforehand would anticipate their economic burden for third countries. Investing in training and capacity building for policymakers and regulators in RIA techniques is essential for effective implementation. Conducting a comparative analysis of the economic impact of EU regulations with alternative regulatory approaches will also help identify the most beneficial options for Guatemala.

---

[35] Edoardo Celeste, 'Digital Constitutionalism, EU Digital Sovereignty Ambitions and the Role of the European Declaration on Digital Rights', REBUILD Centre Working Paper No. 16 (2024) 3.
[36] Edoardo Celeste (n 21) 195.
[37] OECD, *Regulatory Impact Assessment* (OECD 2020) 8.

## 4. International Regulatory Co-operation to implement Collaborative Standard Settings.

International Regulatory Cooperation[38] is an agreement between countries to promote cooperation in the design, monitoring, enforcement, or *ex-post* management of policies and regulations.  The EU should engage Guatemala in discussions and processes on digital policies. This can be accomplished by working together on round tables and projects, exchanging knowledge, and advocating for inclusive standards. These collaborative policies will be tailored to address Guatemala's specific needs while minimizing the possibility of imposing onerous regulations.

## 5. Recognition of Diversity to guarantee inclusive digital policies.

The EU should recognize and respect the diversity of cultural, social, and economic contexts in Guatemala when formulating digital policies. It should involve stakeholders' perspectives to ensure that policies are inclusive and relevant to Guatemala's unique situation. To avoid applying 'the law of the strongest'[39] blindly and mitigate adverse effects, recognition of diversity and international collaboration are essential to ensuring effective EU policies globally.

---

[38] OECD, *International Regulatory Co-Operation* (OECD 2021) 26.
[39] Edoardo Celeste (n 21) 196.

## Conclusion

By adopting these principles, the EU can demonstrate responsible leadership in shaping global digital policies. Neglecting to implement solutions could perpetuate the EU's digital policies as 'rules for everyone written in Brussels'.[40]  Nonetheless, succeeding in their implementation can create a more equitable and sustainable digital future for all.

---

[40] Wall Street Journal Editorial, 'Regulatory Imperialism' (26 October 2007) 1.